\documentstyle[amssymb,12pt]{article}

\newlength{\minitwocolumn}
\font\teneufm=eufm10
\font\seveneufm=eufm7
\font\fiveeufm=eufm5
\newfam\eufmfam
\textfont\eufmfam=\teneufm
\scriptfont\eufmfam=\seveneufm
\scriptscriptfont\eufmfam=\fiveeufm

\makeatletter
\@addtoreset{equation}{section}
\makeatother

\newtheorem{thm}{Theorem}[section]
\newtheorem{prop}[thm]{Proposition}

\newtheorem{conj}[thm]{Conjecture}

\newtheorem{df}{Definition}[section]

\title{\bf The Intergals of Motion for\\
the Deformed $W$-Algebra $W_{q,t}(\widehat{sl_N})$}
\begin{document}

~\\
\begin{center}
\begin{huge}
{\bf The Intergals of Motion for\\
the Deformed $W$-Algebra $W_{q,t}(\widehat{sl_N})$}
\end{huge}
~\\~\\~\\~\\
{ Boris FEIGIN$~^{\alpha}$,
Takeo KOJIMA$~^{\beta}$,
Jun'ichi SHIRAISHI$~^{\gamma}$,
Hidekazu WATANABE$~^{\gamma}$}
\\~\\
{\it
$~^\alpha$
L.D.Landau Institute for Theoretical Sciences, \\
Chernogolovka, Moscow 142432, RUSSIA
\\
$~^\beta$
Department of Mathematics,
College of Science and Technology,
Nihon University,\\
Surugadai, Chiyoda-ku, Tokyo 101-0062, 
JAPAN\\
$~^\gamma$
Graduate School of Mathematical Science,
University of Tokyo, \\
Komaba, Meguro-ku, Tokyo, 153-8914,
JAPAN
}
\end{center}
~\\
\begin{abstract}
We review the deformed $W$-algebra $W_{q,t}(\widehat{sl_N})$
and its screening currents.
We explicitly
construct the Local Integrals of Motion ${\cal I}_n~(n=1,2,\cdots)$
for this deformed $W$-algebra.
We explicitly
construct the Nonlocal Integrals of Motion ${\cal G}_n~(n=1,2,\cdots)$
by means of the screening currents.
Our Integrals of Motion 
commute with each other, 
and give the elliptic version of those
for the Virasoro algebra
and the $W$-algebra $W(\widehat{sl_3})$ 
\cite{BLZ, BHK}.
\end{abstract}

\newpage

\section{Introduction}

V.Bazhanov, S.Lukyanov and Al.Zamolodchikov \cite{BLZ}
constructed field theoretical analogue of 
the commuting transfer matrix ${\bf T}(z)$
as the trace of the free field realization of the monodromy
matrix associated with the quantum affine symmetry 
$U_q(\widehat{sl_2})$.
They call the coefficients of the asymptotic expansion of the
operator ${\rm log}{\bf T}(z)$ at $z \to \infty$,
the Local Integrals of Motioin for the Virasoro algebra.
The Local Integrals of Motion $I_n$ have the form
\begin{eqnarray}
I_{2k-1}=\int_0^{2\pi}\frac{du}{2\pi}T_{2k}(u),~~~(k=1,2,\cdots).
\end{eqnarray}
Here the densities $T_{2k}(u)$ are differential polynomials of 
the energy momentum tensor 
of the Virasoro algebra $T(u)=-\frac{C_{CFT}}{24}+\sum_{n=-\infty}^\infty
L_{-n}e^{\sqrt{-1}nu}$,
where the operators $L_n$
satisfies the commutation relation,
$[L_n,L_m]=(n-m)L_{n+m}+\frac{C_{CFT}}{12}(n^3-n)\delta_{n+m,0}$.
The first few densities $T_{2k}(u)$ are written by
\begin{eqnarray}
T_2(u)=T(u),~
T_4(u)=:T^2(u):,~
T_6(u)=:T^3(u):+\frac{C_{CFT}+2}{12}:(T'(u))^2:.
\end{eqnarray}
However they did not give explicit formulae of general densities $T_{2k}$,
all densities $T_{2k}(u)$ are uniequely
determined by requirement of the commutativity,
$[I_{2k-1},I_{2l-1}]=0,~(k,l=1,2,\cdots)$.
They call the coefficients of the Taylor expnsion of the
operator ${\bf T}(z)$ at $z=0$, 
the Nonlocal Integrals of Motion.
They gave explicit formulae for 
the Nonlocal
Integrals of Motion ${G}_k,~(k=1,2,\cdots)$, 
by means of the screening currents $F_1(z), F_2(z)$.
\begin{eqnarray}
G_k&=&
\int\cdots \int_{2\pi \geqq u_1 \geqq u_2 \geqq \cdots \geqq u_{2k}\geqq 0}
(e^{2\pi \sqrt{-1}P}F_1(u_1)F_2(u_2)F_1(u_3)F_2(u_4)\cdots F_2(u_{2k})\nonumber\\
&+&
e^{-2\pi \sqrt{-1}P}F_2(u_1)F_1(u_2)F_2(u_3)F_1(u_4)\cdots F_1(u_{2k}))
du_1 du_2 \cdots du_{2k},
\end{eqnarray}
where $P$ is zero-mode operator.
Because of the commutativity of the transfer matrix ${\bf T}(z)$,
the integtrals of motion commute with each other :
$[I_{2k-1},I_{2l-1}]=[I_{2k-1},G_l]=[G_k,G_l]=0$.
V.Bazhanov, A.Hibberd and S.Khoroshkin \cite{BHK} 
considered 
the $W$-algebra $W(\widehat{sl_3})$ version of \cite{BLZ}.

V.Bazhanov, S.Lukyanov and Al. Zamolodchikov's
theory \cite{BLZ} can be regarded as the quantum version
of the KdV problem,
as it reduces to the classical KdV problem in classical limit
$C_{CFT}\to -\infty$, under the substitution
\begin{eqnarray}
T(u)\to -\frac{C_{CFT}}{6}U(u),~~[~,~]\to \frac{6\pi}{\sqrt{-1}C_{CFT}}\{~,~\}.
\end{eqnarray}
The Poisson bracket 
structure $\{,\}$ gives the second Hamiltonian structure of 
the KdV equation.
The Local Integrals of Motion $I_{2k-1}$ tend to
$I_{2k-1}^{(cl)}$.
\begin{eqnarray}
I_{2k-1}^{(cl)}=\int_0^{2 \pi}\frac{du}{2\pi}T_{2k}^{(cl)}(u),~~(k=1,2,\cdots),
\end{eqnarray}
where the first few densities are written by
\begin{eqnarray}
T_2^{(cl)}=U(u),~~
T_4^{(cl)}=U^2(u),~~
T_6^{(cl)}=U^3(u)-\frac{1}{2}(U'(u))^2.
\end{eqnarray}
The KdV hierarchy are given by
\begin{eqnarray}
\partial_{t_{2k-1}}U=\{I_{2k-1}^{(cl)},U\},~(k=1,2,\cdots).
\end{eqnarray}


The purpose of this paper is
to construct
the elliptic version of 
the Integrals of Motion $I_{k}$ and $G_k$ 
given by V.Bazhanov, S.Lukyanov and Al.Zamolodchikov \cite{BLZ}
and V.Bazhanov, A.Hibberd and S.Khoroshkin \cite{BHK}.
Our method of construction is completely different from
those of \cite{BLZ, BHK}.
Instead of considering the transfer matrix ${\bf T}(z)$,
we directly give 
explicit formulae of both Local Integrals of Motion
${\cal I}_n$ and the Nonlocal Integrals of Motion ${\cal G}_m$,
for the deformed $W$-algebra $W_{q,t}(\widehat{sl_N})$.
\begin{eqnarray}
[{\cal I}_m,{\cal I}_n]=0,~~
[{\cal G}_m,{\cal G}_n]=0,~~[{\cal I}_m,{\cal G}_n]=0,~~(m,n=1,2,\cdots).
\end{eqnarray}

The organization of this paper is as follows.
In Section 2, we give basic definition,
including bosons, screening currents.
In Section 3, we review the deformed $W$-algebra $W_{q,t}(\widehat{sl_N})$. 
In Section 4, we construct explicit formulae
for the Local Integrals of Motion
${\cal I}_n,~(n=1,2,\cdots)$ for the deformed $W$-algebra 
$W_{q,t}(\widehat{sl_N})$.
In Section 5, we construct explicit formulae
for the Nonlocal Integrals of Motion
${\cal G}_m,~(m=1,2,\cdots)$.

\section{Basic Definition}

In this section we give the basic definition.
Let us fix three parameters 
$0<x<1, r \in {\mathbb C}$ and $s \in {\mathbb C}$.

\subsection{Bosons}

Let $\epsilon_i (1\leqq i \leqq N)$
be an orthonormal basis in ${\mathbb R}^N$
relative to the standard basis in ${\mathbb R}^N$
relative to the standard inner product $(,)$.
Let us set $\bar{\epsilon}_i=\epsilon_i-\epsilon,
\epsilon=\frac{1}{N}\sum_{j=1}^N \epsilon_j$.
We idetify $\epsilon_{N+1}=\epsilon_1$.
Let $P=\sum_{i=1}^N {\mathbb Z}\bar{\epsilon}_i$ the weight lattice.
Let us set $\alpha_i=\bar{\epsilon}_i-\bar{\epsilon}_{i+1}\in P$.

Let $\beta_m^j$ be the oscillators
$(1\leqq j \leqq N, m \in {\mathbb Z}-\{0\})$
with the commutation relations
\begin{eqnarray}
~[\beta_m^i,\beta_n^j]=\left\{
\begin{array}{cc}
m\frac{[(r-1)m]}{[r m]}\frac{[(s-1)m]}{[s m]}\delta_{n+m,0}&
(1\leqq i=j\leqq N)\\
-m\frac{[(r-1)m]}{[r m]}\frac{[m]}{[sm]}x^{sm~{\rm sgn}(i-j)}
\delta_{n+m,0}
&(1\leqq i\neq j \leqq N)
\end{array}
\right.
\end{eqnarray}
Here the symbol $[a]$ stands for $\frac{x^a-x^{-a}}{x-x^{-1}}$.

We also introduce the zero mode operator $P_\lambda$, $(\lambda \in P)$.
They are ${\mathbb Z}$-linear in $\lambda$
and satisfy
\begin{eqnarray}
[iP_\lambda, Q_\mu]=(\lambda,\mu),~~(\lambda,\mu \in P).
\end{eqnarray}

Let us intrduce the bosonic Fock space ${\cal F}_{l,k} (l,k \in P)$
generated by $\beta_{-m}^j (m>0)$ over the vacuum vector $|l,k \rangle$ :
\begin{eqnarray}
{\cal F}_{l,k}={\mathbb C}[
\{\beta_{-1}^j,\beta_{-2}^j,\cdots\}_{1\leqq j \leqq N}]|l,k\rangle,
\end{eqnarray}
where
\begin{eqnarray}
\beta_m^j|l,k\rangle&=&0,  (m>0),\\
P_\alpha |l,k\rangle&=&
\left(\alpha, \sqrt{\frac{r}{r-1}}l-\sqrt{\frac{r-1}{r}}k\right)|l,k\rangle,\\
|l,k\rangle&=&e^{i\sqrt{\frac{r}{r-1}}Q_l-i\sqrt{\frac{r-1}{r}}Q_k}|0,0\rangle.
\end{eqnarray}

Let us set the Dynkin-diagram automorphism $\eta$ by
\begin{eqnarray}
\eta(\beta_m^1)=x^{-\frac{2s}{N}m}\beta_m^2,\cdots,
\eta(\beta_m^{N-1})=x^{-\frac{2s}{N}m}\beta_m^N,
~\eta(\beta_m^N)=x^{\frac{2s}{N}(N-1)m}\beta_m^1,
\end{eqnarray}
and
$\eta(\epsilon_i)=\epsilon_{i+1},~(1\leqq i \leqq N)$.

\subsection{Basic Operators}

In this section we introduce the basic operators.
Let us set $z=x^{2u}$.

\begin{df}~~~We set the screening currents $F_j(z) (1\leqq j \leqq N)$ by
\begin{eqnarray}
F_j(z)&=&e^{i\sqrt{\frac{r-1}{r}}Q_{\alpha_j}}
(x^{(\frac{2s}{N}-1)j}z)^{\sqrt{\frac{r-1}{r}}P_{\alpha_j}+\frac{r-1}{r}}
\nonumber\\
&\times&
:\exp\left(\sum_{m\neq 0}\frac{1}{m}B_m^jz^{-m}\right):,~(1\leqq j\leqq N-1)\\
F_N(z)&=&e^{i\sqrt{\frac{r-1}{r}}Q_{\alpha_N}}
(x^{2s-N}z)^{\sqrt{\frac{r-1}{r}}P_{\bar{\epsilon}_N}+\frac{r-1}{2r}}
(z)^{-\sqrt{\frac{r-1}{r}}P_{\bar{\epsilon}_1}+\frac{r-1}{2r}}\nonumber\\
&\times&
:\exp\left(
\sum_{m\neq 0}\frac{1}{m}B_m^N
z^{-m}
\right):.
\end{eqnarray}
Here we set
\begin{eqnarray}
B_m^j&=&(\beta_m^j-\beta_m^{j+1})x^{-\frac{2s}{N}
jm},~(1\leqq j \leqq N-1),\\
B_m^N&=&(x^{-2sm}\beta_m^N-\beta_m^1).
\end{eqnarray}
\end{df}
The screening currents $F_j(z),~(1\leqq j \leqq N-1)$ are studied well in
\cite{FJMOP}. We introduce new current $F_N(z)$, which can be regarded
as ``affinization'' of screenings $F_j(z),~~(1\leqq j \leqq N-1)$.

In what follows, the symbol $[u]_r$ stands for the theta function satisfying
\begin{eqnarray}
~[u+r]_r&=&-[u]_r=[-u]_r,\\
~[u+\tau]_r&=&-e^{\frac{2\pi i}{r}(u+\frac{\tau}{2})}[u]_r,~~
{\rm where}~\tau=\frac{\pi i}{{\rm log}x}.
\end{eqnarray}
Explicitly it is given by
\begin{eqnarray}
[u]_r&=&x^{u^2/r-u}\Theta_{x^{2r}}(x^{2u}),\\
\Theta_q(z)&=&(z;q)_\infty (q/z;q)_\infty (q;q)_\infty,\\
(z;q)_\infty&=&\prod_{j=0}^\infty (1-zq^j).
\end{eqnarray}

\begin{prop}~~~
The screening currents $F_j(z),~(1\leqq j \leqq N)$ satisfy
the following commutation relations,
\begin{eqnarray}
\frac{1}{[u_1-u_2-\frac{s}{N}+1]_r}F_j(z_1)F_{j+1}(z_2)
&=&\frac{1}{[u_2-u_1+\frac{s}{N}]_r}
F_{j+1}(z_2)F_{j}(z_1),~~(1\leqq j \leqq N),\nonumber\\\\
\frac{[u_1-u_2]_r}{[u_1-u_2-1]_r}F_j(z_1)F_{j}(z_2)&=&
\frac{[u_2-u_1]_r}{
[u_2-u_1-1]_r}F_{j}(z_2)F_{j}(z_1),~~(1\leqq j \leqq N),
\end{eqnarray}
and
\begin{eqnarray}
F_i(z_1)F_j(z_2)&=&F_j(z_2)F_i(z_1),~~(|i-j|\geqq 2).
\end{eqnarray}
We read $F_{N+1}(z)=F_1(z)$.
\end{prop}

\begin{prop}~~~
The action of $\eta$ on the screenings $F_j(z)$ is given by
\begin{eqnarray}
&&\eta(F_j(z))=F_{j+1}(x^{1-\frac{2s}{N}}z),~~(1\leqq j \leqq N-2),\\
&&\eta(F_{N-1}(z))=F_N(x^{1-\frac{2s}{N}}z)x^{(2s-N)(-\sqrt{\frac{r-1}{r}}
P_{\bar{\epsilon}_1}+\frac{r-1}{r})},\\
&&\eta(F_{N}(z))=F_1(x^{1-\frac{2s}{N}}z)x^{(2s-N)(\sqrt{\frac{r-1}{r}}
P_{\bar{\epsilon}_1}+\frac{r-1}{r})}.
\end{eqnarray}
We have
$\eta(F_{N-1}(z_1)F_N(z_2))=F_N(z_1)F_1(z_2)$.
\end{prop}

\begin{df}~~We set the fundamental operator 
$\Lambda_j(z), (1\leqq j \leqq N)$ by
\begin{eqnarray}
\Lambda_j(z)&=&x^{-2\sqrt{r(r-1)} P_{\bar{\epsilon}_j}}
:\exp
\left(\sum_{m \neq 0} \frac{x^{rm}-x^{-rm}}{m}
\beta_m^j z^{-m}\right):~~(1\leqq j \leqq N).
\end{eqnarray}
\end{df}

\begin{prop}~~
The action of $\eta$ on the screenings $\Lambda_j(z)$ is given by
\begin{eqnarray}
\eta(\Lambda_j(z))=\Lambda_{j+1}(xz),~~(1\leqq j \leqq N-1),~~~
\eta(\Lambda_N(z))=\Lambda_1(x^{1-2s}z).
\end{eqnarray}
\end{prop}

\begin{prop}~~The screening currents $F_j(z),~(1\leqq j\leqq N)$
and the fundamental operators 
$\Lambda_j(z),~(1\leqq j \leqq N)$ commute up to delta-function
$\delta(z)=\sum_{n\in {\mathbb Z}}z^m$. 
\begin{eqnarray}
~[\Lambda_j(z_1),F_j(z_2)]&=&(-1+x^{-2r+2})\delta
\left(x^{\frac{2s}{N}j-r}\frac{z_2}{z_1}\right)
:\Lambda_j(z_1)F_j(z_2):,~~(1\leqq j \leqq N),\nonumber
\\
\\
~[\Lambda_{j+1}(z_1),F_j(z_2)]&=&(1-x^{-2r+2})\delta
\left(x^{\frac{2s}{N}j+r}\frac{z_2}{z_1}\right)
:\Lambda_{j+1}(z_1)F_j(z_2):,~~(1\leqq j \leqq N-1),\nonumber\\
\\
~[\Lambda_1(z_1),F_N(z_2)]&=&(1-x^{-2r+2})
\delta\left(x^{r}\frac{z_2}{z_1}\right)
:\Lambda_1(z_1)F_N(z_2):.
\end{eqnarray}
\end{prop}


\section{Deformed $W$-Algebra $W_{q,t}(\widehat{sl_N})$}

In this section we review the deformed $W$-algebra $W_{q,t}(\widehat{sl_N})$
\cite{SKAO, AKOS, Odake}. 

\subsection{Deformation of $W$-Algebra}

In this section we give the deformation
of the $W$-algebra.

\begin{df}
Let us set the operator $T_j(z),~(1\leqq j \leqq N)$ by
\begin{eqnarray}
T_j(z)=\sum_{1\leqq s_1<s_2<\cdots<s_j\leqq N}
:\Lambda_{s_1}(x^{-j+1}z)\Lambda_{s_2}(x^{-j+3}z)\cdots \Lambda_{s_j}(x^{j-1}z):.
\end{eqnarray}
\end{df}

\begin{prop}~~The bosonic operators $T_j(z),~(1\leqq j \leqq N)$ satisfy
the following relations.
\begin{eqnarray}
&&f_{i,j}(z_2/z_1)T_i(z_1)T_j(z_2)
-f_{j,i}(z_1/z_2)T_j(z_2)T_i(z_1)\nonumber\\
&=&c
\sum_{k=1}^i
\prod_{l=1}^{k-1}\Delta(x^{2l+1})\times
\left(\delta\left(\frac{x^{j-i+2k}z_2}{z_1}\right)
f_{i-k,j+k}(x^{-j+i})T_{i-k}(x^{-k}z_1)T_{j+k}(x^kz_2)\right.\nonumber\\
&-&\left.
\delta\left(\frac{x^{-j+i-2k}z_2}{z_1}\right)
f_{i-k,j+k}(x^{j-i})T_{i-k}(x^{k}z_1)T_{j+k}(x^{-k}z_2)
\right),~~(1\leqq i \leqq j \leqq N),\label{def:3parameter}
\end{eqnarray}
where $\delta(z)=\sum_{n \in {\mathbb Z}}z^n$.\\
Here we set the constnt $c$ and the auxiliary function $\Delta(z)$ by
\begin{eqnarray}
c=-\frac{(1-x^{2r})(1-x^{-2r+2})}{(1-x^2)},~~
\Delta(z)=\frac{(1-x^{2r-1}z)(1-x^{1-2r}z)}{(1-xz)(1-x^{-1}z)}.
\label{def:Delta}
\end{eqnarray}
Here we set the structure functions,
\begin{eqnarray}
f_{i,j}(z)=\exp\left(
\sum_{m=1}^\infty
\frac{1}{m}(1-x^{2rm})(1-x^{-2(r-1)m})
\frac{(1-x^{2m Min(i,j)})(1-x^{2m(s-Max(i,j))})}
{(1-x^{2m})(1-x^{2sm})}x^{|i-j|m}z^m
\right).\nonumber\\
\end{eqnarray}
\end{prop}
{\bf Example}~~For $N=2$ the operators $T_1(z), T_2(z)$ satisfy
\begin{eqnarray}
&&f_{1,1}(z_2/z_1)T_1(z_1)T_1(z_2)-
f_{1,1}(z_1/z_2)T_1(z_2)T_1(z_1)\nonumber\\
&=&c(\delta(x^2z_2/z_1)T_2(x z_2)-
\delta(x^2z_1/z_2)T_2(x^{-1}z_2)),\\
&&f_{1,2}(z_2/z_1)T_1(z_1)T_2(z_2)=
f_{2,1}(z_1/z_2)T_2(z_2)T_1(z_1),\\
&&f_{2,2}(z_2/z_1)T_2(z_1)T_2(z_2)=
f_{2,2}(z_1/z_2)T_2(z_2)T_2(z_1).
\end{eqnarray}
{\bf Example}
~~For $N=3$ the operators $T_1(z), T_2(z), T_3(z)$ satisfy
\begin{eqnarray}
&&f_{1,1}(z_2/z_1)T_1(z_1)T_1(z_2)-
f_{1,1}(z_1/z_2)T_1(z_2)T_1(z_1)\nonumber\\
&=&c(\delta(x^2z_2/z_1)T_2(x z_2)-
\delta(x^2z_1/z_2)T_2(x^{-1}z_2)),\\
&&f_{1,2}(z_2/z_1)T_1(z_1)T_2(z_2)-
f_{2,1}(z_1/z_2)T_2(z_2)T_1(z_1)\nonumber\\
&=&c(\delta(x^3 z_2/z_1)T_3(x z_2)-
\delta(x^3 z_1/z_2)T_3(x^{-1}z_2)),\\
&&f_{2,2}(z_2/z_1)T_2(z_1)T_2(z_2)-
f_{2,2}(z_1/z_2)T_2(z_2)T_2(z_1)\nonumber\\
&=&c f_{1,3}(1)(\delta(x^2 z_2/z_1)T_1(xz_2)T_3(x z_2)-
\delta(x^2 z_1/z_2)T_1(x^{-1}z_2)T_3(x^{-1}z_2)),\\
&&f_{1,3}(z_2/z_1)T_1(z_1)T_3(z_2)=
f_{3,1}(z_1/z_2)T_3(z_2)T_1(z_1),\\
&&f_{2,3}(z_2/z_1)T_2(z_1)T_3(z_2)=
f_{3,2}(z_1/z_2)T_3(z_2)T_2(z_1),\\
&&f_{3,3}(z_2/z_1)T_3(z_1)T_3(z_2)=
f_{3,3}(z_1/z_2)T_3(z_2)T_3(z_1).
\end{eqnarray}

\begin{df}~~
The three parameter deformed $W$-algebra 
is an associative algebra generated by $T_n^{(j)},~(n \in {\mathbb Z},
1\leqq j \leqq N)$. Defining relations are given by (\ref{def:3parameter}).
The elements $T_n^{(j)}$ are Fourier coefficients of
$T_j(z)=\sum_{n\in {\mathbb Z}}T_n^{(j)}z^{-n}$.
\end{df}

For general $N$, 
upon specialization $s=N$, the operator
$T_N(z)$ degenerate to scalar $1$.
Let us set parameters $q=x^{2r}$ and $t=x^{2r-2}$.

\begin{thm}~~Upon specialization $s=N$, the operator $T_j(z)~(1\leqq j \leqq N)$
give a free field realization of the deformed $W$-algebra 
$W_{q,t}(\widehat{sl_N})$.
\end{thm}

~\\
{\bf Example}~~For $N=2$ and $s=2$, 
the operator $T_2(z)$ degenerates to scalar $1$, 
and Fourier coefficients of $T_1(z)=\sum_{n \in {\mathbb Z}}T_n z^{-n}$
satisfy the defining relation of
the deformed Virasoro algebra ${Vir}_{q,t}=
W_{q,t}(\widehat{sl_2})$ \cite{SKAO}.
\begin{eqnarray}
[T_n,T_m]=-\sum_{l=0}^\infty f_l(T_{n-l}T_{m+l}-T_{m-l}T_{n+l})
+c((q/t)^n-(t/q)^n)\delta_{n+m,0},
\end{eqnarray}
where we set the structure constant $f_l$ by
$f_{11}(z)=1+\sum_{l=1}^\infty f_l z^l$.
Let us set $q=e^h$ and $t=q^\beta$, $(\beta=\frac{r-1}{r})$,
and take the limit $h\to 0$ under the following
$h$-expansion,
\begin{eqnarray}
T_n=2\delta_{n,0}+\beta\left(L_n+\frac{(1-\beta)^2}{4\beta}\delta_{n,0}\right)h^2
+O(h^4),
\end{eqnarray}
we have the defining relation of the Virasoro algebra,
\begin{eqnarray}
[L_m,L_n]=(n-m)L_{m+n}+\frac{C_{CFT}}{12}n(n^2-1)\delta_{n+m,0},~~
C_{CFT}=1-\frac{6(1-\beta)^2}{\beta}.
\end{eqnarray}

\subsection{Comparsion with another definition}

At first glance, our definition
of the deformed $W$-algebra is different from
those in \cite{AKOS, Odake}.
In this section we show they are essentially the same thing.
Let us set the element ${\cal C}_m$ by
\begin{eqnarray}
{\cal C}_m=\sum_{j=1}^N x^{(N-2j+1)m}\beta_m^j.
\end{eqnarray}
This element ${\cal C}_m$ is $\eta$-invariant,
$\eta({\cal C}_m)={\cal C}_m$.
Let us divide $\Lambda_j(z)$ into $\Lambda_j^{DWA}(z)$ and ${\cal Z}(z)$.
\begin{eqnarray}
\Lambda_j(z)&=&\Lambda_j^{DWA}(z){\cal Z}(z),
~~(1\leqq j \leqq N),
\end{eqnarray}
where we set
\begin{eqnarray}
\Lambda_j^{DWA}(z)&=&
x^{-2\sqrt{r(r-1)} P_{\bar{\epsilon}_j}}
:\exp
\left(\sum_{m \neq 0} \frac{x^{rm}-x^{-rm}}{m}
\left(\beta_m^j-\frac{[m]_x}{[Nm]_x}{\cal C}_m
 \right)z^{-m}\right):,\\
{\cal Z}(z)&=&:\exp\left(\sum_{m\neq 0}\frac{x^{rm}-x^{-rm}}{m}
\frac{[m]_x}{[Nm]_x}{\cal C}_m z^{-m}\right):.
\end{eqnarray}
Let us set 
\begin{eqnarray}
T_j^{DWA}(z)=\sum_{1\leqq s_1<s_2<\cdots <s_j\leqq N}
:\Lambda_{s_1}^{DWA}(x^{-j+1}z)\Lambda_{s_2}^{DWA}(x^{-j+3}z)
\cdots \Lambda_{s_j}^{DWA}(x^{j-1}z):.
\end{eqnarray}

\begin{thm}~~~The operators $T_j^{DWA}(z),~(1\leqq j \leqq N-1)$ give a free field
realization of the deformed $W$-algebra $W_{q,t}(\widehat{sl_N})$
\cite{AKOS, Odake}.
\end{thm}

\begin{prop}~~~
The operators $T_j^{DWA}(z)$ and ${\cal Z}(z)$ commutes with each other.
\begin{eqnarray}
T_j^{DWA}(z_1){\cal Z}(z_2)={\cal Z}(z_2)T_j^{DWA}(z_1),~~~
(1\leqq j \leqq N-1).
\end{eqnarray}
\end{prop}

Three paramedter deformed $W$-algebra defined in the previous subsection
(\ref{def:3parameter}),
can be regarded 
as an extension of $W_{q,t}(\widehat{sl_N})$,
therefore, we sometime call three parameter deformed $W$-algebra,
``the deformed $W$-algebra $W_{q,t}(\widehat{sl_N})$''.
In what follows, mainly,  we consider three parameter deformed $W$-algebra.
It is simpler to show the commutation relations of the Integrals of
Motion
$[{\cal I}_m,{\cal I}_n]=[{\cal G}_m,{\cal G}_n]=
[{\cal I}_m,{\cal G}_n]=0$ for three parameter $(x,r,s)$ deformed case than
those for two parameter $(x,r,s=N)$ deformed case.
One additionnal parameter $s$ resolves singularity
in the Integrals of Motion, and make problem simpler.

\section{Local Integrals of Motion}

In this section we give
explicit formulae of the Local Integrals of Motion ${\cal I}_n$.

In what follows we use the notation of the ordered product.
\begin{eqnarray}
\prod_{\longrightarrow
\atop{l\in L}}T_j(z_l)=T_j(z_{l_1})T_j(z_{l_2})\cdots T_j(z_{l_m}),~~
(L=\{l_1,l_2,\cdots,l_m|l_1<l_2<\cdots<l_m\}).
\end{eqnarray}
For formal power series ${\cal A}(z_1,z_2,\cdots,z_n)=
\sum_{k_1,k_2,\cdots,k_n \in {\mathbb Z}}a_{k_1,k_2,\cdots,k_n}
z_1^{k_1} z_2^{k_2} \cdots z_n^{k_n}$, we set the symbol 
$[\cdots]_{1, z_1, z_2, \cdots, z_n}$.
\begin{eqnarray}
[{\cal A}(z_1,z_2,\cdots,z_n)]_{1,z_1,z_2,\cdots,z_n}=a_{0,0,\cdots,0}.
\end{eqnarray}

Let us set the auxiliary function $g_{i,j}(z)$ by fusion of
$g_{1,1}(z)=f_{1,1}(z)$.
\begin{eqnarray}
g_{i,1}(z)=g_{1,1}(x^{-i+1}z)g_{1,1}(x^{-i+3}z)\cdots g_{1,1}(x^{i-1}z),
\nonumber\\
g_{i,j}(z)=g_{i,1}(x^{-j+1}z)g_{i,1}(x^{-j+3}z)\cdots g_{i,1}(x^{j-1}z).
\end{eqnarray}

\begin{df}~~We set the operator ${\cal O}_n(z_1,z_2,\cdots,z_n)$ by
\begin{eqnarray}
&&{\cal O}_n(z_1,z_2,\cdots,z_n)=
\sum_{\alpha_1,\alpha_2,\alpha_3,\cdots,\alpha_N\geqq 0
\atop{\alpha_1+2\alpha_2+3\alpha_3+\cdots+N\alpha_N=n}}
\sum_{A_1^{(1)},\cdots,A_{\alpha_1}^{(1)},
\cdots, A_1^{(N)},\cdots,A_{\alpha_N}^{(N)}\subset
\{1,2,\cdots,n\}
\atop{
|A_j^{(t)}|=t,~~
\oplus_{j,t}A_j^{(t)}=\{1,2,\cdots,n\}}}\nonumber\\
&\times&
\prod_{\longrightarrow
\atop{j \in A_{Min}^{(1)}}}T_1(z_j)
\prod_{\longrightarrow
\atop{j \in A_{Min}^{(2)}}}T_2(x^{-1}z_j)
\cdots
\prod_{\longrightarrow
\atop{j \in A_{Min}^{(t)}}}T_t(x^{-1+t-2[\frac{t}{2}]}z_j)
\cdots
\prod_{\longrightarrow
\atop{j \in A_{Min}^{(N)}}}T_N(x^{-1+N-2[\frac{N}{2}]}z_j)\nonumber\\
&\times&
\prod_{t=1}^N
\left((-c)^{t-1}
\prod_{u=1}^{t-1}
\Delta(x^{2u+1})^{t-u-1}\right)^{\alpha_t}
\prod_{t=1}^N
\prod_{j=1
\atop{j_1=A_{j,1}^{(t)}
\atop{\cdots
\atop{j_t=A_{j,t}^{(t)}}
}}}^{\alpha_t}
\sum_{\sigma \in S_t
\atop{\sigma(1)=1}}
\prod_{u=1
\atop{u \neq [\frac{t}{2}]+1}}^t
\delta\left(\frac{x^2z_{j_{\sigma(u+1)}}}{z_{j_{\sigma(u)}}}\right)
\nonumber\\
&\times&
\prod_{t=1}^N \prod_{j<k
\atop{j,k \in A_{Min}^{(t)}}}g_{t,t}\left(\frac{z_k}{z_j}\right)
\prod_{1\leqq t<u \leqq N}\prod_{j\in A_{Min}^{(t)}
\atop{k\in A_{Min}^{(u)}}}g_{t,u}\left(
x^{u-t-2[\frac{u}{2}]+2[\frac{t}{2}]}\frac{z_k}{z_j}
\right).
\end{eqnarray}
Here we have set the constant $c$ and the function
$\Delta(z)$ in (\ref{def:Delta}).
When the index set $A_j^{(t)}=
\{j_1,j_2,\cdots, j_t|j_1<j_2<\cdots<j_t\}$,
$(1\leqq t \leqq N, 1\leqq j \leqq \alpha_t)$,
we set $A_{j,k}^{(t)}=j_k$, and
$A_{Min}^{(t)}=\{A_{1,1}^{(t)},A_{2,1}^{(t)},
\cdots, A_{t,1}^{(t)}\}$.
\end{df}
{\bf Example}
\begin{eqnarray}
{\cal O}_1(z)&=&T_1(z),\\
{\cal O}_2(z_1,z_2)&=&g_{1,1}(z_2/z_1)T_1(z_1)T_1(z_2)-
c\delta(x^2z_2/z_1)T_2(x^{-1}z_1),\\
{\cal O}_3(z_1,z_2,z_3)&=&
g_{11}(z_2/z_1)g_{1,1}(z_3/z_1)g_{1,1}(z_3/z_2)T_1(z_1)T_1(z_2)T_1(z_3)\nonumber\\
&-&cg_{1,2}(x^{-1}z_2/z_1)T_1(z_1)
\delta(x^2z_3/z_2)T_2(x^{-1}z_2)\nonumber\\
&-&cg_{1,2}(x^{-1}z_1/z_2)
T_1(z_2)\delta(x^2z_3/z_1)T_2(x^{-1}z_1)\nonumber\\
&-&c
g_{1,2}(x^{-1}z_1/z_3)T_1(z_3)
\delta(x^2z_2/z_1)T_2(x^{-1}z_1)\nonumber\\
&+&c^2\Delta(x^3)
(\delta(x^2z_2/z_1)\delta(x^2z_1/z_3)
+\delta(x^2z_1/z_2)\delta(x^2z_3/z_1))T_3(z_1).
\end{eqnarray}

Let us set the auxiliary function $s(z)=s(1/z)$ by
\begin{eqnarray}
s(z)=\frac{(z;x^{2s})_\infty (x^{2s-2r}z;x^{2s})_\infty}
{(x^{2s-2}z;x^{2s})_\infty (x^{-2r+2}z;x^{2s})_\infty}
\times 
\frac{(1/z;x^{2s})_\infty (x^{2s-2r}/z;x^{2s})_\infty}
{(x^{2s-2}/z;x^{2s})_\infty (x^{-2r+2}/z;x^{2s})_\infty}.
\end{eqnarray}

~\\
\begin{df}~~~For ${\rm Re}(s)>0$ and ${\rm Re}(r)<0$,
we define a family of the operators ${\cal I}_n,
~(n=1,2,\cdots)$ by
\begin{eqnarray}
{\cal I}_n&=&\left[\prod_{1\leqq j<k \leqq n}s(z_k/z_j)
{\cal O}_n(z_1,\cdots,z_n)\right]_{1,z_1,\cdots,z_n}.
\end{eqnarray} 
For generic ${\rm Re}(s)>0$ and $r \in {\mathbb C}$,
the definition of ${\cal I}_n$ should be understood as analytic continuation.
We call the operator ${\cal I}_n$ the Local Integral Motion for
the deformed $W$-algebra $W_{q,t}(\widehat{sl_N})$.
\end{df}

\begin{prop}~~
The operator ${\cal O}_n(z_1,z_2,\cdots,z_n)$ is $S_n$-invariant 
in ``weakly sense''.
\begin{eqnarray}
\prod_{1\leqq j<k \leqq n}s(z_k/z_j){\cal O}_n(z_1,z_2,\cdots,z_n)=
\prod_{1\leqq j<k \leqq n}s(z_{\sigma(k)}/z_{\sigma(j)})
{\cal O}_n(z_{\sigma(1)},z_{\sigma(2)},\cdots,z_{\sigma(n)}),~
(\sigma \in S_n).\nonumber\\
\end{eqnarray}
\end{prop}
The following is one of {\bf Main Results}.

\begin{thm}~~~The Local Integrals of Motion ${\cal I}_n~(n=1,2,\cdots)$
commute with each other.
\begin{eqnarray}
~[{\cal I}_m, {\cal I}_n]=0,~~(m,n=1,2,\cdots).
\end{eqnarray}
\end{thm}

~\\
By using $S_n$-invariance of ${\cal O}_n(z_1,z_2,\cdots,z_n)$ in
``weakly sense'',
the above theorem is reduced to the following theta function identity,
which is shown by induction \cite{FO}.
\begin{eqnarray}
&&\sum_{J \subset \{1,2,\cdots,n+m\}
\atop{|J|=n}}
\prod_{j \in J}
\prod_{k \notin J}
\frac{[u_k-u_j+1]_s[u_k-u_j+r-1]_s}{[u_{k}-u_{j}]_s[u_k-u_j+r]_s},
\nonumber\\
&=&\sum_{J^c \subset \{1,2,\cdots,n+m\}
\atop{|J^c|=m}}
\prod_{j \in J^c}
\prod_{k \notin J^c}
\frac{[u_k-u_j+1]_s[u_k-u_j+r-1]_s}{[u_{k}-u_{j}]_s[u_k-u_j+r]_s}.
\end{eqnarray}

\begin{conj}~~The Local Integrals of Motion are $\eta$-invariant.
\begin{eqnarray}
\eta({\cal I}_n)={\cal I}_n,~~~(n=1,2,\cdots).
\end{eqnarray}
\end{conj}

We have checked for small $n$.
We have already shown $\eta$-invariance conjecture,
$\eta({\cal I}_n)={\cal I}_n$, 
for the deformed Virasoro algebra case, 
$Vir_{q,t}=W_{q,t}(\widehat{sl_2})$ \cite{FKSW}.

\section{Nonlocal Integrals of Motion}

In this section we give explicit formulae of
the Nonlocal Integrals of Motion.
Let us set the theta function
$\vartheta(u^{(1)}|u^{(2)}|\cdots|u^{(N)})$ by following conditions.
\begin{eqnarray}
&&
\vartheta(u^{(1)}|\cdots|u^{(t)}+r|\cdots|u^{(N)})
=\vartheta(u^{(1)}|\cdots|u^{(t)}|\cdots|u^{(N)}),~~(1\leqq t \leqq N)\\
&&
\vartheta(u^{(1)}|\cdots|u^{(t)}+r\tau|\cdots|u^{(N)})\nonumber\\
&=&
e^{-2\pi i \tau-\frac{2\pi i}{r}(u_{t-1}-2u_t+u_{t+1}+\sqrt{r(r-1)}P_{\alpha_t})}
\vartheta(u^{(1)}|\cdots|u^{(t)}|\cdots|u^{(N)}),~~(1\leqq t \leqq N),\\
&&\vartheta(u^{(1)}+k|\cdots|u^{(N)}+k)=
\vartheta(u^{(1)}|\cdots|u^{(N)}),~~(k\in {\mathbb C}),\\
&&\eta(\vartheta(u^{(1)}|\cdots|u^{(N)}))=
\vartheta(u^{(N)}|u^{(1)}|\cdots|u^{(N-1)}).
\end{eqnarray}

~\\
{\bf Example}~~For $N=2$ case, we have
\begin{eqnarray}
\vartheta(u_1|u_2)&=&[u_1-u_2-\sqrt{r(r-1)}P_{\alpha_1}+\alpha]_r
[u_1-u_2-\alpha]_r\nonumber\\
&+&[u_1-u_2-\sqrt{r(r-1)}P_{\alpha_1}-\alpha]_r
[u_1-u_2+\alpha]_r,~~(\alpha \in {\mathbb C}).
\end{eqnarray}

\begin{df}~~~For ${\rm Re}(r)\neq 0$ and $0<{\rm Re}(s)<2$,
we define a family of operators ${\cal G}_m,~(m=1,2,\cdots)$ by
\begin{eqnarray}
{\cal G}_m&=&\prod_{t=1}^N \prod_{j=1}^m 
\oint_C \frac{dz_j^{(t)}}{2\pi \sqrt{-1}z_j^{(t)}}
F_1(z_1^{(1)})\cdots F_1(z_m^{(1)})
F_2(z_1^{(2)})\cdots F_2(z_m^{(2)})\cdots
F_N(z_1^{(N)})\cdots F_N(z_m^{(N)})\nonumber\\
&\times&
\frac{\displaystyle
\prod_{t=1}^N \prod_{1\leqq i<j \leqq m}
\left[u_i^{(t)}-u_j^{(t)}\right]_r
\left[u_j^{(t)}-u_i^{(t)}-1\right]_r
}{
\displaystyle
\prod_{t=1}^{N-1}\prod_{i,j=1}^m 
\left[u_i^{(t)}-u_j^{(t+1)}+1-\frac{s}{N}\right]_r
\prod_{i,j=1}^m 
\left[u_i^{(1)}-u_j^{(N)}+\frac{s}{N}\right]_r}\nonumber\\
&\times&
\vartheta\left(\sum_{j=1}^m u_j^{(1)}\right|
\left.\sum_{j=1}^m u_j^{(2)}\right|
\cdots
\left|\sum_{j=1}^m u_j^{(N)}\right).
\end{eqnarray}
Here the integral contour $C$ is given by
\begin{eqnarray}
&&|x^{\frac{2s}{N}}z_j^{(t+1)}|
<|z_i^{(t)}|<|x^{-2+\frac{2s}{N}}z_j^{(t+1)}|,~~(1\leqq t \leqq N-1, 1\leqq i,j \leqq m),\\
&&
|x^{2-\frac{2s}{N}}z_j^{(1)}|<|z_i^{(N)}|<|x^{-\frac{2s}{N}}z_j^{(1)}|,
~~(1\leqq i,j \leqq m).
\end{eqnarray}
For generic $s \in {\mathbb C}$, the definition of ${\cal G}_n$
should be understood as analytic continuation.
We call the operator ${\cal G}_n$ the Nonlocal Integrals of Motion
for the deformed $W$-algebra $W_{q,t}(\widehat{sl_N})$.
\end{df}

~\\
{\bf Example}~~For $N=2$ and $m=1$ case, we have
\begin{eqnarray}
{\cal G}_1=
\int \int_C \frac{dz_1}{2\pi \sqrt{-1}z_1}
\frac{dz_2}{2\pi \sqrt{-1}z_2}
F_1(z_1)F_0(z_2)\frac{\vartheta(u_1|u_2)}{[u_1-u_2+\frac{s}{2}]_r
[u_1-u_2-\frac{s}{2}+1]_r}.
\end{eqnarray}
Here $C$ is given by
$|x^{s}z_2|<|z_1|<|x^{-2+s}z_2|$.

~\\
The following is one of {\bf Main Results}.

\begin{thm}~~The Nonlocal Integrals of Motion ${\cal G}_n,~(n=1,2,\cdots)$
commute with each other.
\begin{eqnarray}
[{\cal G}_m,{\cal G}_n]=0,~~~(m,n=1,2,\cdots).
\end{eqnarray}
\end{thm}

~\\

By using commutation relations of the screening currents
$F_j(z)$,
the above theorem is reduced to the following theta function identity,
which is shown by induction \cite{FO}.

\begin{eqnarray}
&&\sum_{\sigma_1 \in S_{m+n}}
\sum_{\sigma_2 \in S_{m+n}}
\cdots \sum_{\sigma_N \in S_{m+n}}
\vartheta_\alpha
\left(\sum_{j=1}^m u_{\sigma_1(j)}^{(1)}\right|
\left.\sum_{j=1}^m u_{\sigma_2(j)}^{(2)}\right|
\cdots
\left|\sum_{j=1}^m u_{\sigma_N(j)}^{(N)}\right)\nonumber\\
&\times&
\vartheta_\beta
\left(\sum_{j=m+1}^{m+n} u_{\sigma_1(j)}^{(1)}\right|
\left.\sum_{j=m+1}^{m+n} u_{\sigma_2(j)}^{(2)}\right|
\cdots
\left|\sum_{j=m+1}^{m+n} u_{\sigma_N(j)}^{(N)}\right)
\nonumber\\
&\times&
\prod_{t=1}^N
\prod_{i=1}^m
\prod_{j=m+1}^{m+n}
\frac{\left[u_{\sigma_t(i)}^{(t)}-u_{\sigma_{t+1}(j)}^{(t+1)}-\frac{s}{N}\right]_r
\left[u_{\sigma_t(j)}^{(t)}-u_{\sigma_{t+1}(i)}^{(t+1)}+1-\frac{s}{N}\right]_r
}{
\left[u_{\sigma_t(i)}^{(t)}-u_{\sigma_{t}(j)}^{(t)}\right]_r
\left[u_{\sigma_t(j)}^{(t)}-u_{\sigma_{t}(i)}^{(t)}-1\right]_r
}\nonumber
\\
&=&
\sum_{\sigma_1 \in S_{m+n}}
\sum_{\sigma_2 \in S_{m+n}}
\cdots \sum_{\sigma_N \in S_{m+n}}
\vartheta_\beta
\left(\sum_{j=1}^n u_{\sigma_1(j)}^{(1)}\right|
\left.\sum_{j=1}^n u_{\sigma_2(j)}^{(2)}\right|
\cdots
\left|\sum_{j=1}^n u_{\sigma_N(j)}^{(N)}\right)
\nonumber\\
&\times&
\vartheta_\alpha
\left(\sum_{j=n+1}^{m+n} u_{\sigma_1(j)}^{(1)}\right|
\left.\sum_{j=n+1}^{m+n} u_{\sigma_2(j)}^{(2)}\right|
\cdots
\left|\sum_{j=n+1}^{m+n} u_{\sigma_N(j)}^{(N)}\right)
\nonumber\\
&\times&
\prod_{t=1}^N
\prod_{i=1}^n
\prod_{j=n+1}^{m+n}
\frac{\left[u_{\sigma_t(i)}^{(t)}-u_{\sigma_{t+1}(j)}^{(t+1)}-\frac{s}{N}\right]_r
\left[u_{\sigma_t(j)}^{(t)}-u_{\sigma_{t+1}(i)}^{(t+1)}+1-\frac{s}{N}\right]_r
}{
\left[u_{\sigma_t(i)}^{(t)}-u_{\sigma_{t}(j)}^{(t)}\right]_r
\left[u_{\sigma_t(j)}^{(t)}-u_{\sigma_{t}(i)}^{(t)}-1\right]_r
},
\end{eqnarray}
where $\vartheta_\alpha(u^{(1)}|u^{(2)}|\cdots|u^{(N)})$ and 
$\vartheta_\beta(u^{(1)}|u^{(2)}|\cdots|u^{(N)})$ 
are not necesarry the same
theta functions.

~\\

\begin{thm}~~The Nonlocal Integrals of Motion are $\eta$-invariant.
\begin{eqnarray}
\eta({\cal G}_m)={\cal G}_m,~~~(m=1,2,\cdots).
\end{eqnarray}
\end{thm}

\begin{conj}~~~The Local Integrals of Motion ${\cal I}_n,~(n=1,2,\cdots)$
and Nonlocal Integrals of Motion ${\cal G}_m,~(m=1,2,\cdots)$
commute with each other.
\begin{eqnarray}
[{\cal I}_n,{\cal G}_m]=0,~~~(m,n=1,2,\cdots).
\end{eqnarray}
\end{conj}

We have already shown
the commutation relations
$[{\cal I}_n,{\cal G}_m]=0, (m,n=1,2,\cdots)$
for the deformed Virasoro algebra case, $Vir_{q,t}=
W_{q,t}(\widehat{sl_2})$ \cite{FKSW}.
We have already shown 
the commutation relations
$[{\cal I}_1,{\cal G}_m]=0,~(m=1,2,\cdots)$
for general $W_{q,t}(\widehat{sl_N})$ case.
If we assume $\eta$-invariance of the Local Integrals of Motion
$\eta({\cal I}_n)={\cal I}_n$,
the commutation relation $[{\cal I}_n,{\cal G}_m]=0$,
$(m,n=1,2,\cdots)$ for general $W_{q,t}(\widehat{sl_N})$, 
can be shown by simple argument.

~\\
{\bf Acknowledgements.}~~We would like to thank Professor M.Jimbo
for his interest to this work.
B.F.is partly supported by Grant RFBR (05-01-01007),
SS (2044.2003.2), and RFBR-JSPS (05-01-02934).
T.K. is partly supported by Grant-in Aid for 
Young Scientist {\bf B} (18740092) from JSPS. 
J.S. is partly supported by Grant-in Aid for 
Scientific Research {\bf C} (16540183) from JSPS.

~\\
\begin{center}
{\it Proceedings for Representation Theory} 2006,
Atami, Japan, p.102-114 (2006).
\end{center}
~~~~~~~~[ISBN4-9902328-2-8]

\end{document}